# Optimizing Active Cyber Defense

Wenlian Lu[1,2], Shouhuai Xu[3], and Xinlei Yi[1]

[1] School of Mathematical Sciences, Fudan University
Shanghai, P. R. China, 200433
Emails: {`wenlian,11210180008`}`@fudan.edu.cn`
[2] Department of Computer Science, University of Warwick
Coventry CV4 7AL, UK
[3] Department of Computer Science, University of Texas at San Antonio
San Antonio, Texas 78249, USA
Email: `shxu@cs.utsa.edu`

**Abstract.** Active cyber defense is one important defensive method for combating cyber attacks. Unlike traditional defensive methods such as firewall-based filtering and anti-malware tools, active cyber defense is based on spreading "white" or "benign" worms to combat against the attackers' malwares (i.e., malicious worms) that also spread over the network. In this paper, we initiate the study of *optimal* active cyber defense in the setting of strategic attackers and/or strategic defenders. Specifically, we investigate infinite-time horizon optimal control and fast optimal control for strategic defenders (who want to minimize their cost) against non-strategic attackers (who do not consider the issue of cost). We also investigate the Nash equilibria for strategic defenders and attackers. We discuss the cyber security meanings/implications of the theoretic results. Our study brings interesting open problems for future research.

## 1 Introduction

The importance of cyber security is well recognized now. However, our understanding of cyber security is still at its infant stage. In general, the attackers are constantly escalating their attack power and sophistication, while the defenders largely lag behind. To be specific, we mention the following *asymmetry* between cyber attack and cyber defense: The effect of malware-like attacks is *automatically amplified* by the network connectivity, while the defense effect is not. This phenomenon had been implied by many previous results (e.g., [28, 9, 6, 26, 34]), but was not explicitly pointed out until very recently [35]. The asymmetry is fundamentally caused by that the defense is *reactive*, including intrusion detection systems, firewalls and anti-malware tools. The asymmetry can be eliminated by the idea of *active cyber defense* [35], where the defender also aims to take advantage of the network connectivity. The concept of active cyber defense is not completely new because researchers have proposed for years the idea of using the defender's "white" or "benign" worms to combat against the attackers' malwares

[5, 1, 29, 23, 16, 18, 13, 30]. In a sense, active cyber defense already happened in practice; for example, the `Welchia` worm attempted to "kill" the Blaster worm in compromised computers [23, 20]. It appears that full-fledged active cyber defense is perhaps inevitable in the near future according to some recent reports [18, 24, 31]. It is therefore more imperative than ever to systematically characterize the effectiveness of active cyber defense. This motivates the present study.

### 1.1 Our Contributions

This paper is inspired by the recent mathematical model of active cyber defense dynamics [35], which characterizes the effect of various model parameters (including the underlying complex network structures) in the setting where *neither the attacker nor the defender is strategic* (i.e., both the attacker and the defender do not consider the issue of cost). Here we study a new perspective of active cyber defense, namely the strategic interaction between the attacker and the defender. On one hand, our study moves a step beyond [35] because we incorporate control-theoretic and game-theoretic models to accommodate strategic interactions. On the other hand, our study assumes away the underlying complex network structures that are explicitly investigated in [35]. This means that our study is essentially based on the *homogeneous* (or *well-mixed*) assumption that each compromised computer can attack the same portion of computers. Tackling the problem of strategic attack-defense interactions with explicit complex network structures is left for future research. Therefore, we deem the present paper as a significant first step toward ultimately understanding the effectiveness of strategic active cyber defense. Specifically, we make the following contributions.

First, we investigate two flavors of optimal control for strategic defenders against non-strategic attackers: *infinite-time horizon* optimal control and *fast* optimal control. In the setting of infinite-time horizon optimal control for the defender, we characterize the conditions under which the defender should adjust its active cyber defense power in a certain quantitative fashion. For example, we identify a condition under which the defender should give up using active cyber defense alone, and instead should resort to other defense methods as well (e.g., proactive defense). In the setting of fast optimal control, where the defender wants to occupy a certain portion of the network as soon as possible and at the minimal cost, there is a significant difference between the case that the active defense cost is linear and the case that the active defense cost is quadratic.

Second, we identify the Nash equilibrium strategies when both the defender and the attacker are strategic. The findings are interesting. For example, when the defender (or attacker) is reluctant to use/expose its advanced active cyber defense tools (or zero-day exploits), it will give up escalating its active defense (or attack) power; otherwise, there are three scenarios: (i) If the defender (or attacker) initially occupies only a certain small portion of the network, it will give up escalating its active defense (or attack). (ii) If the defender (or attacker) initially occupies a certain significant portion of the network, it will escalate its active defense (or attack) as much as possible. (iii) If the defender (or attacker) initially occupies a certain large portion of the network, it will not escalate its active defense (or attack) — a sort of diminishing returns.

The rest of the paper is structured as follows. Section 2 briefly reviews the related prior work. Section 3 describes the basic active cyber defense model under the homogeneous assumption. Section 4 investigates optimal control for strategic defenders against non-strategic attackers. Section 5 studies Nash equilibria for strategic defenders and attackers. Section 6 concludes the paper with some open problems. Lengthy proofs are deferred to the Appendix. The main notations used in the paper are listed below:

| | |
|---:|:---|
| $\alpha_B, \alpha_R$ | defender **B**'s defense power $\alpha_B$ and attacker **R**'s attack power $\alpha_R$ |
| $i_B(t), i_R(t)$ | portions of the nodes occupied respectively by the defender and the attacker at time $t$, where $i_B(t) + i_R(t) = 1$ |
| $\pi_B, \pi_B(t)$ | $\pi_B$ is control variable and $\pi_B(t)$ is control function |
| $\hat{\pi}_B$ | solution in the infinite-time horizon optimal control case |
| $\pi_B^*, \pi_B^{**}$ | solutions in the case of fast optimal control with linear and quadratic cost functions, respectively |
| $z$ | discount rate |
| $k_B$ | normalization ratio between the defender's detection cost and recovery cost |
| $\lambda$ | normalization ratio between the unit of time and the defender's active defense cost |
| $k_R$ | normalization ratio between the attacker's maintenance cost and penetration cost |

## 2 Related Work

Our investigation is built on recent studies in *mathematical computer malware models*. These models originated in the *mathematical biological epidemic models* introduced in the 1920's [19, 12], which were first adapted to study the spreading of computer virus in the 1990's [10, 11]. All these models made the homogeneous assumption that each individual (e.g., computer) in the population has equally infection effect on the other individuals in the population, and the assumption that the infected individuals recover because of reactive defense (e.g., anti-malware tools). In the past decade, there were many studies that aim to eliminate the aforementioned homogeneous assumption, by explicitly incorporating the heterogeneous network structures [28, 9, 6, 26, 34, 32]. The mathematical tools used for these studies are *Dynamical Systems* in nature. These studies demonstrated that the attack effect of malware spreading against reactive defense is automatically amplified by the largest eigenvalue of the adjacency matrix, which represents the underlying complex network structure. This is the attack-defense asymmetry phenomenon mentioned above.

The attack-defense asymmetry phenomenon motivated the study of mathematical models of *active cyber defense* [35], which is a relatively new sub-field in cyber security [18, 24, 31] as previous explorations were mainly geared toward legal and policy issues [5, 1, 29, 23, 16, 18, 13, 30]. One real-life incident of the flavor of active cyber defense is that the `Welchia` worm attempted to "kick out" another kind of worms (e.g., the Blaster worm) [23, 20]. In the first mathematical characterization of active cyber defense [35], neither the attacker nor the defender is strategic (i.e., they do not consider the issue of cost), albeit the model accommodates the underlying complex network structure. In the present paper, we move a step toward ultimately understanding *optimal* active cyber defense, where the attacker and/or the defender are/is strategic (i.e., they want to minimize their cost). Finally, we note that automatic patching [27] is not active cyber defense because automatic patching aims to *prevent* attacks, whereas active cyber defense aims to identify and possibly clean up infected computers.

There have been many studies (e.g., [33, 21, 8, 4, 14, 22, 15, 25]) on applying Control Theory and Game Theory to understand various issues related to computer malware spreading. Our study is somewhat inspired by the botnet-defense model investigated in [4]. All the studies mentioned above only considered *reactive* defense; whereas we investigate how to optimize *active* cyber defense. For general information about the applications of Control Theory and Game Theory to cyber security, we refer to [2, 17] and the references therein.

## 3 The Basic Active Cyber Defense Model

Consider a population of nodes, which can abstract computers in a cyber system. At any point in time, a node is either occupied by defender **B** (i.e., the node is secure), or occupied by attacker **R** (i.e., the node is compromised). Denote by $i_B(t)$ the portion of nodes that are occupied by the defender at time $t$, and by $i_R(t)$ the portion of nodes that are occupied by the attacker at time $t$, where $i_B(t)+i_R(t) = 1$ for any $t \geq 0$. In the interaction between cyber attack and active cyber defense, the defender and the attacker can "grab" nodes from each other in the same fashion. Let $\alpha_B$ abstract defender **B**'s power in grabbing attacker-occupied nodes using active cyber defense, and $\alpha_R$ abstract attacker **R**'s power in compromising defender-occupied nodes using malware-like cyber attacks. Under the homogeneous assumption that (i) each secure node has the same power in "grabbing" the attacker-occupied nodes and (ii) each compromised node has the same power in compromising the defender-occupied nodes, we obtain the following Dynamical System model:

$$\begin{cases} \frac{di_B(t)}{dt} = \alpha_B i_B(t) i_R(t) - \alpha_R i_R(t) i_B(t) \\ \frac{di_R(t)}{dt} = \alpha_R i_R(t) i_B(t) - \alpha_B i_B(t) i_R(t), \end{cases}$$

where $i_B(t) + i_R(t) = 1$, $i_B(t) \geq 0$, and $i_R(t) \geq 0$ for all $t \geq 0$. Due to the symmetry, we only need to consider

$$\frac{di_B(t)}{dt} = \alpha_B i_B(t)(1 - i_B(t)) - \alpha_R i_B(t)(1 - i_B(t)). \tag{1}$$

If neither the attacker nor the defender is strategic (i.e., they do not consider the issue of cost), the dynamics of system (1) can be characterized as follows.

- If the attacker is more powerful than the defender, namely $\alpha_R > \alpha_B$, the attacker will occupy the entire network in the fashion of the Logistic equation (i.e., when $i_R$ is small, $i_R$ increases slowly; when $i_R$ is around a threshold value, $i_R$ increases exponentially; when $i_R$ is large, $i_R$ increases slowly).
- If the defender is more powerful than the attacker, namely $\alpha_B > \alpha_R$, the defender will occupy the network in the same fashion as in the above case.
- If the attacker and the defender are equally powerful, namely $\alpha_R = \alpha_B$, the system state is in equilibrium. In other words, $i_B(t) = i_B(0)$ and $i_R(t) = i_R(0) = 1 - i_B(0)$ for any $t > 0$.

The above model accommodates non-strategic attackers and non-strategic defenders, and is the starting point for our study of optimal active cyber defense.

## 4 Optimal Control for Strategic Defender Against Non-Strategic Attacker

### 4.1 Infinite-time Horizon Optimal Control

In this setting, the non-strategic attacker **R** maintains a *fixed* degree of attack power $\alpha_R$, while the defender **B** is strategic. That is, the strategic defender aims to minimize its cost (specified below) by adjusting its defense power $\alpha_B$ via

$$\alpha_B = b + \pi_B(a - b),$$

while obeying the dynamics of (1), where $\pi_B \in [0, 1]$ is the control variable and $\alpha_B \in [b, a]$ is the defender's defense power with $1 \geq a > b \geq 0$. The cost to the defender consists of two parts.

- The *recovery cost* for recovering the compromised nodes to secure states (e.g., re-installing the operating systems and updating the backup data files, interference with the computers' routine functions). We represent this cost by $f_B(i_B(t))$ for some real-valued function $f_B(\cdot)$. We assume $f'_B(\cdot) < 0$ because the more nodes the defender occupies, the lower the cost for the defender to recover the compromised nodes.
- The *detection cost* for detecting (or recognizing) compromised nodes via active cyber defense, which partly depends on the attack's evasiveness. We represent this cost by $k_B \cdot \pi_B(\cdot)$, where $k_B$ is the normalization ratio between the detection cost and the recovery cost, and $\pi_B(\cdot)$ is the control function that specifies the adjustable degree of active cyber defense power. This is plausible because using more powerful active defense mechanisms (e.g., more sophisticated/advanced "white" worms) causes a higher cost but allows the defender to fight against the attacks more effectively.

The above definition of cost accommodates at least the following family of active cyber defense: The defender uses "white" worms to detect the compromised nodes, then possibly manually recovers the compromised nodes. This is perhaps the most probable scenario because for example, the attacker's malware may have corrupted or deleted some data files in the compromised computers. Note that the detection cost highlights the difference between (i) active-cyber-defense based detection, where the defender's detection tools (i.e., "white" worms) do not reside on the compromised computers, and (ii) reactive-cyber-defense based detection such as the current generation of anti-virus software, where the detection tools do not spread over the network.

Assuming that the attacker maintains a fixed degree of attack power $\alpha_R$, the defender's *optimization goal* is to minimize the total cost with a constant discount rate $z$ over an infinite-time horizon, namely

$$\inf_{0 \leq \pi_B(\cdot) \leq 1} \left\{ J_B(\pi_B(\cdot)) = \int_0^\infty e^{-zt}(f_B(i_B(t)) + k_B \cdot \pi_B(t))dt \right\}, \quad (2)$$

where $f'_B(\cdot) < 0$, $\pi_B(\cdot) \in [0, 1]$, and the attacker's fixed degree of attack power $\alpha_R$ is treated as a constant. Now the optimization problem reduces to identifying the optimal defense strategy $\hat{\pi}_B$. To solve the minimization problem, we use Pontryagin's Minimum Principle to find the Hamiltonian associated to (2):

$$\begin{aligned} &H_B(i_B, \pi_B, p) \\ &= f_B(i_B) + k_B \pi_B + p[\alpha_B i_B(1 - i_B) - \alpha_R i_B(1 - i_B)] \\ &= (k_B + p i_B(1 - i_B)(a - b))\pi_B + f_B(i_B) + p b i_B(1 - i_B) - p \alpha_R i_B(1 - i_B) \end{aligned} \quad (3)$$

where $p$ is the adjoint equation

$$\begin{cases} \dot{p} &= -\frac{\partial H_B}{\partial i_B} + zp = -f'_B(i_B) + p[z - (\alpha_B - \alpha_R)(1 - 2i_B)] \\ p_1(\infty) &= 0. \end{cases} \quad (4)$$

The *optimal strategy* $\hat{\pi}_B$ is obtained by minimizing the Hamiltonian $H_B(i_B, \pi_B, p)$. Since $H_B(i_B, \pi_B, p)$ is linear in $\pi_B$, the optimal control strategy $\hat{\pi}_B$ takes the following bang-bang control form:

$$\hat{\pi}_B = \begin{cases} 1 & if \ \frac{\partial H_B}{\partial \pi_B} < 0 \\ u_B \ (0 < u_B < 1, \text{ to be determined}) & if \ \frac{\partial H_B}{\partial \pi_B} = 0 \\ 0 & if \ \frac{\partial H_B}{\partial \pi_B} > 0 \end{cases} \quad (5)$$

where $\frac{\partial H_B}{\partial \pi_B} = k_B + p i_B(1 - i_B)(a - b)$. In the singular form $\frac{\partial H_B}{\partial \pi_B} = 0$ and for a period of time, we have

$$p = \frac{-k_B}{i_B(1 - i_B)(a - b)}. \quad (6)$$

Further differentiating $\frac{\partial H_B}{\partial \pi_B}$ with respect to $t$, we have

$$\frac{d}{dt}\left(\frac{\partial H_B}{\partial \pi_B}\right) = \dot{p}i_B(1-i_B)(a-b) + p(1-2i_B)\dot{i}_B(a-b)$$

$$= i_B(1-i_B)(a-b)\left\{-f'_B(i_B) + p[z-(\alpha_B-\alpha_R)(1-2i_B)]\right\}$$

$$+ p(1-2i_B)(a-b)\left\{\alpha_B i_B(1-i_B) - \alpha_R i_B(1-i_B)\right\}$$

$$= -i_B(1-i_B)(a-b)f'_B(i_B) - k_B z$$

Define $F_B(i_B) = -i_B(1-i_B)(a-b)f'_B(i_B) - k_B z$. Then we need to study the roots of $F_B(\cdot) = 0$.

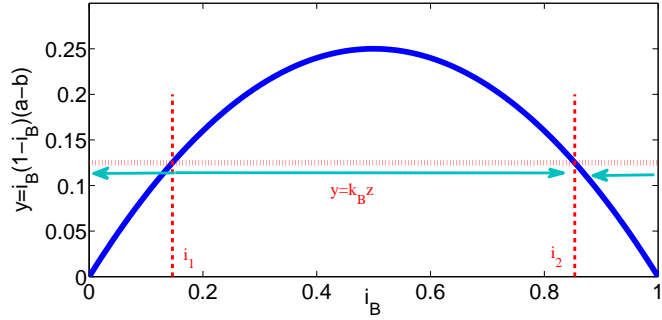

**Fig. 1.** Illustration of the roots of $F_B(i_B) = 0$ with $f_B(i_B) = 1 - i_B$, $a-b = 1$ and $k_B z = 1/8$, where the $x$-axis represents $i_B$ and the $y$-axis represents $y(i_B) = i_B(1-i_B)(a-b)$. The arrows indicate the directions the outcome under optimal control will head for.

Before presenting the results, we discuss the ideas behind them. In this paper, we focus on case $f_B(i_B) = 1 - i_B$, which can be easily extended to any linear recovery-cost function. If $k_B z < \frac{1}{4}(a-b)$, then $F_B(i_B) = 0$ has two roots:

$$i_1 = \frac{1 - \sqrt{1 - 4\frac{k_B z}{a-b}}}{2} \quad \text{and} \quad i_2 = \frac{1 + \sqrt{1 - 4\frac{k_B z}{a-b}}}{2}$$

with $0 < i_1 < i_2 < 1$. As illustrated in Figure 1, this implies

$$\begin{cases} F_B(i_B) < 0 & if\ i_B < i_1 \\ F_B(i_B) > 0 & if\ i_1 < i_B < i_2 \\ F_B(i_B) < 0 & if\ i_B > i_2. \end{cases}$$

Then, the optimal strategy $\hat{\pi}_B$ of the singular form can be obtained by solving $\ddot{i}_B\mid_{i_B=i_1\ or\ i_B=i_2} = 0$.

**Theorem 1.** *Suppose the non-strategic attacker maintains a fixed degree of attack power $\alpha_R$, $f_B(i_B) = 1 - i_B$ and $k_B z < \frac{1}{4}(a-b)$. Let $i_1 < i_2$ be the roots of $F_B(i_B) = 0$. Let $u_B = \frac{\alpha_R - b}{a - b}$. The optimal control strategy for defender **B** is:*

$$\hat{\pi}_B = \begin{cases} 0 & if\ i_B < i_1 \\ u_B & if\ i_B = i_1 \\ 1 & if\ i_1 < i_B < i_2 \\ u_B & if\ i_B = i_2 \\ 0 & if\ i_B > i_2 \end{cases} \tag{7}$$

Proof of Theorem 1 is deferred to Appendix A. In practice, $i_1$ and $i_2$ can be obtained numerically. Theorem 1 (also as illustrated in Figure 1) shows that the outcome of the infinite-time horizon optimal control, namely $\lim_{t \to \infty} i_B(t)$, depends on the initial system state $i_B(0)$ as follows:

- If $1 > i_B(0) > i_2$, the defender should use the least powerful/costly active defense mechanisms (i.e., $\alpha_B = b$) because $\hat{\pi}_B = 0$. Moreover, the outcome of the optimal defense is that the defender will occupy $i_2$ portion of the network, namely $\lim_{t \to \infty} i_B(t) = i_2$. This suggests a sort of *diminishing returns* in active cyber defense: It is more cost-effective to pursue "good enough" security (i.e., $\lim_{t \to \infty} i_B(t) = i_2 < 1$) than to pursue "perfect" security (i.e., $\lim_{t \to \infty} i_B(t) = 1$) even if it is possible.
- If $0 = i_B(0) < i_1$, the defender should use the least powerful/costly active defense mechanisms (i.e., $\alpha_B = b$) because $\hat{\pi}_B = 0$. Moreover, the outcome of the optimal defense is that the defender should give up (using active cyber defense as the only defense methods), as the attacker will occupy the entire network, namely $\lim_{t \to \infty} i_B(t) = 0$. In other words, the defender should resort to other defense methods as well (e.g., proactive defense).
- If $i_B(0) \in (i_1, i_2)$, the defender should use the most powerful/costly active defense mechanisms (i.e., $\alpha_B = a$) because $\hat{\pi}_B = 1$. Moreover, the outcome of the optimal defense is that the defender will occupy $i_2$ portion of the network, namely $\lim_{t \to \infty} i_B(t) = i_2$. This also suggests a sort of *diminishing returns* mentioned above.
- If $i_B(0) = i_1$ or $i_B(0) = i_2$, the defender should adjust its deployment of active cyber defense mechanisms according to $u_B = \frac{\alpha_R - b}{a - b}$, which means $\alpha_B = \alpha_R$. Moreover, the outcome of the optimal defense is that $i_B(t) = i_B(0)$ for all $t > 0$.

Now we consider the degenerated scenarios of $k_B z \geq 1/4(a-b)$. The proof is similar to, but much simpler than, the proof of Theorem 1, and thus omitted.

**Theorem 2.** *Suppose the non-strategic attacker maintains a fixed degree of attack power $\alpha_R$ and $f_B(i_B) = 1 - i_B$.*

- *If $k_B z = 1/4(a-b)$, then $F_B(i_B) = 0$ has only one root, $i_1 = i_2 = \frac{1}{2}$. The optimal control strategy is*

$$\hat{\pi}_B = \begin{cases} 0 & if\ i_B < i_1 \\ u_B = \frac{\alpha_R - b}{a - b} & if\ i_B = i_1 \\ 0 & if\ i_B > i_1. \end{cases} \tag{8}$$

- If $k_B z > 1/4(a - b)$, then $F_B(i_B) = 0$ has no root. The optimal control strategy is $\hat{\pi}_B = 0$.

The *cyber security implications* of Theorem 2 are the following. In the case $k_B z = \frac{1}{4}(a - b)$, the outcome under the optimal control depends on the initial system state as follows:

- If $1 > i_B(0) > i_1$, the defender should use the least powerful/costly active cyber defense mechanisms because $\hat{\pi}_B = 0$. The outcome is that the defender will occupy $i_1$ portion of the network, namely $\lim_{t \to \infty} i_B(t) = i_1$.
- If $0 = i_B(0) < i_1$, the defender should use the least powerful/costly active cyber defense mechanisms because $\hat{\pi}_B = 0$. The outcome is that the defender will give up using active cyber defense alone, as the attacker will occupy the entire the network, namely $\lim_{t \to \infty} i_B(t) = 0$. In other words, the defender should resort to other defense methods as well (e.g., proactive defense).
- If $i_B(0) = i_1$, the defender will adjust its degree of active cyber defense power according to $\hat{\pi}_B = u_B = \frac{\alpha_R - b}{a - b}$, which means $\alpha_B = \alpha_R$. The outcome is that $i_B(t) = i_B(0)$ for all $t > 0$.

In the case $k_B z > 1/4(a-b)$, the defender should use the least powerful/costly active cyber defense mechanisms because $\hat{\pi}_B = 0$. The outcome is that $\lim_{t \to \infty} i_B(t) = 0$, meaning that the defender should give up using active cyber defense alone and resort to other defense methods as well (e.g., proactive defense).

By considering Theorems 1 and 2 together, we draw some deeper insights. Specifically, for a given $z$, different $k_B$'s suggest different optimal active defense strategies. More specifically, if $k_B > \frac{1}{4z}(a - b)$, meaning that the cost of optimal control is dominating, then defender **B** should use the least powerful/costly active cyber defense mechanisms because $\hat{\pi}_B(t) = 0$ for all $t$ and the outcome is $\lim_{t \to \infty} i_B = 0$. In other words, the defender should give up using active cyber defense alone, and resort to other kinds of defense methods as well (e.g., proactive defense). If $k_B < \frac{1}{4z}(a - b)$, meaning that the cost of control is not dominating, the defender should enforce optimal control according to the initial state $i_B(0)$. In particular, if $k_B = 0$, meaning that the special case that the cost of control is not counted, defender **B** should use the most powerful/costly active defense mechanisms as $\hat{\pi}_B(t) = 1$ for all $t$, and the outcome is that $\lim_{t \to \infty} i_B = 1$, namely that the defender will occupy the entire network.

### 4.2 Fast Optimal Control for Strategic Defenders against Non-Strategic Attackers

Now we consider *fast* optimal control for strategic defenders against non-strategic attackers, as motivated by the following question: Suppose the attacker maintains a fixed degree of attack power $\alpha_R$ and the defender initially occupies $i_B(0) = i_0 < i_e$ portions of the nodes, how can the defender use optimal control to occupy the desired $i_e$ portions of the nodes *as soon as possible*? More precisely, the optimization is to minimize *the sum of active defense cost and time* (after

appropriate normalization), which can be described by the following functional:

$$J_F(\pi_B(\cdot)) = T + \lambda \int_0^T h(\pi_B(t))dt$$

where $h(\cdot)$ is the cost function with respect to the control function $\pi_B(\cdot)$. We consider two scenarios of cost functions: linear and quadratic. In both scenarios, we need to identify defender **B**'s optimal strategy with respect to the dynamics of (1) and a given objective $i_e > i_0$ for some *hitting time T* that is to be identified.

**Scenario I: Fast optimal control with linear cost functions.** In this scenario, we have $h(\pi_B) = \pi_B$. The optimization task is to minimize the active defense cost plus the time $T$:

$$\inf_{0 \le \pi_B(\cdot) \le 1} \left\{ J_F(\pi_B(\cdot)) = T + \lambda \int_0^T \pi_B(t)dt \right\} \quad (9)$$

$$\text{subject to} \begin{cases} \frac{di_B(t)}{dt} = \alpha_B i_B(t)(1 - i_B(t)) - \alpha_R i_B(t)(1 - i_B(t)) \\ i_B(0) = i_0 \\ i_B(T) = i_e \end{cases}$$

where $\lambda > 0$ is the normalization ratio between the unit of time and the active defense cost $\int_0^T \pi_B(t)dt$, and $i_0 < i_e$. That is, $\lambda$, $i_0$ and $i_e$ are given, but $T$ is free. Note that the active defense cost $\int_0^T \pi_B(t)dt$ includes both detection and recovery cost, where $\pi_B(t)$ is the control function.

**Theorem 3.** *The solution to the fast optimal control problem (9) is*

$$(\pi_B^*, T^*) = (1, T_1), \quad (10)$$

*where* $T_1 = \frac{1}{a - \alpha_R} \ln\left(\frac{i_e}{1-i_e} \frac{1-i_0}{i_0}\right)$.

Proof of Theorem 3 is deferred to Appendix B. The *cyber security implication* of Theorem 3 is the following. In order to achieve fast optimal control, the defender should use the most powerful/costly active cyber defense mechanisms, namely $\pi_B(t) = 1$ for $t < T^*$, until the system state becomes $i_B(T^*) = i_e$ at time $T^*$. After time $T^*$, if the defender continues enforcing $\pi_B(t) = 1$ for $t > T^*$, then $\lim_{t \to \infty} i_B(t) = 1$, meaning that the defender will occupy the entire network.

**Scenario II: Fast optimal control with quadratic cost functions.** In this scenario, we have $h(\pi_B) = \pi_B^2$. The optimization task is to minimize the following sum of active defense cost and time, which differs from the linear cost (9) in that the cost function $\pi_B$ is replaced with cost function $\pi_B^2$:

$$\inf_{0 \le \pi_B(\cdot) \le 1} \left\{ J_F(\pi_B(\cdot)) = T + \lambda \int_0^T \pi_B^2(t)dt \right\} \quad (11)$$

$$\text{subject to} \begin{cases} \frac{di_B(t)}{dt} = \alpha_B i_B(t)(1 - i_B(t)) - \alpha_R i_B(t)(1 - i_B(t)) \\ i_B(0) = i_0 \\ i_B(T) = i_e \end{cases}$$

where $\lambda > 0$ is the ratio between the unit of time and the active defense cost $\int_0^T \pi_B^2(t)dt$ (including both recovery cost and detection cost), and $i_0 < i_e$. That is, $\lambda$, $i_0$ and $i_e$ are given, but $T$ is free.

**Theorem 4.** *The solution to the fast optimal control problem (11) is*

$$(\pi_B^{**}, T^{**}) = \begin{cases} (u^*, T_2), \; if \; \lambda \geq \frac{a-b}{a+b-2\alpha_R} \; and \; a - b > 2(\alpha_R - b), \\ (1, T_3), \; otherwise \end{cases} \quad (12)$$

*where*

$$u^* = \frac{\alpha_R - b}{a - b} + \sqrt{\left(\frac{b - \alpha_R}{a - b}\right)^2 + \frac{1}{\lambda}},$$

$$T_2 = \frac{1}{b + (a - b)u^* - \alpha_R} \ln\left(\frac{i_e}{1 - i_e} \frac{1 - i_0}{i_0}\right),$$

$$T_3 = \frac{1}{a - \alpha_R} \ln\left(\frac{i_e}{1 - i_e} \frac{1 - i_0}{i_0}\right).$$

Proof of Theorem 4 is deferred to Appendix C. It *cyber security implication* is: Unlike in the setting of linear cost function (Theorem 3), the defender should not necessarily enforce the most powerful/costly active cyber defense mechanisms as $\pi_B^{**}$ is not always equal to 1. If the defender continues enforcing $\pi_B(t) = 1$ for $t > T^{**}$ after the system reaches state $i_B(T^{**}) = i_e$ at time $T^{**}$, the defender will occupy the entire network, namely $\lim_{t \to \infty} i_B(t) = 1$.

## 5 Nash Equilibria for Strategic Attacker and Defender

Now we ask the question: What if the attacker is also strategic? Analogous to the way of modeling strategic defenders, we assume $\alpha_R \in [b, a]$. (It is straightforward to extend the current setting $\alpha_B, \alpha_R \in [b, a]$ to the setting $\alpha_B \in [b_B, a_B]$ and $\alpha_R \in [b_R, a_R]$.) A strategic attacker can adjust its attack power

$$\alpha_R = b + \pi_R(a - b),$$

via control variable $\pi_R(\cdot) \in [0, 1]$. That is, the attacker can launch more sophisticated attacks (i.e., greater $\pi_R$ leading to greater $\alpha_R$), which however incurs higher cost (e.g., the investment for obtaining more powerful attack tools).

Since both the defender and the attacker are strategic, we naturally consider a game-theoretic model. Specifically, the defender **B**'s optimization task is

$$\phi_B(i_B) = \inf_{0 \leq \pi_B(\cdot) \leq 1} \left\{ J_B(\pi_B(\cdot), \pi_R(\cdot)) = \int_0^\infty e^{-zt}(f_B(i_B(t)) + k_B \cdot \pi_B(i_B(t)))dt \right\},$$

and the attacker **R**'s optimization task is

$$\phi_R(i_B) = \inf_{0 \leq \pi_R(\cdot) \leq 1} \left\{ J_R(\pi_B(\cdot), \pi_R(\cdot)) = \int_0^\infty e^{-zt}(f_R(i_B(t)) + k_R \cdot \pi_R(i_B(t)))dt \right\},$$

where $\pi_B(\cdot), \pi_R(\cdot) \in [0,1]$, $f'_B(\cdot) < 0$ (as in the infinite-time horizon optimal control case investigated above), $f'_R(\cdot) > 0$ because $f_R(i_B(t))$ represents the *maintenance* cost to the attacker, $k_R$ is the normalization ratio between the attacker's maintenance cost and *penetration* cost (which depends on the capability of the attack tools), and $k_R \cdot \pi_R(\cdot)$ is the *penetration* cost. Note that $f'_R(\cdot) > 0$ is relevant because the attacker may need to conduct some costly (or risky) activities after "grabbing" a node from the defender (e.g., downloading attack payloads from some remote server, while this downloading operation may increase the chance that the compromised node is detected by active defense). Since $f'_R(\cdot) > 0$ implies $df_R/di_R < 0$, the attacker's optimization task for $\pi_R$ is in parallel to the optimization for $\pi_B$. The Hamiltonians associated to defender **B**'s and attacker **R**'s optimization problems are:

$$H_B(i_B, \pi_B(i_B), \pi_R(i_B), p_1)$$
$$= f_B(i_B) + k_B \pi_B + p_1[\alpha_B i_B(1-i_B) - \alpha_R i_B(1-i_B)]$$
$$= (k_B + p_1 i_B(1-i_B)(a-b))\pi_B + f_B(i_B) + p_1 b i_B(1-i_B) - p_1 \alpha_R i_B(1-i_B);$$

$$H_R(i_B, \pi_B(i_B), \pi_R(i_B), p_2)$$
$$= f_R(i_B) + k_R \pi_R + p_2[\alpha_B i_B(1-i_B) - \alpha_R i_B(1-i_B)]$$
$$= (k_R - p_2 i_B(1-i_B)(a-b))\pi_R + f_R(i_B) + p_2 \alpha_B i_B(1-i_B) - p_2 b i_B(1-i_B).$$

The adjoint equation is

$$\begin{cases} \dot{p}_1 = -\frac{\partial H_B}{\partial i_B} + z p_1 = -f'_B(i_B) + p_1[z - (\alpha_B - \alpha_R)(1 - 2i_B)] \\ p_1(\infty) = 0 \\ \dot{p}_2 = -\frac{\partial H_R}{\partial i_B} + z p_2 = -f'_R(i_B) + p_2[z - (\alpha_B - \alpha_R)(1 - 2i_B)] \\ p_2(\infty) = 0. \end{cases}$$

**Theorem 5.** *Suppose $f_B(i_B) = 1 - i_B$, $f_R(i_B) = i_B$. Then, the Nash equilibria under various scenarios are listed in Table 1, where $F_B(i_B) = -i_B(1-i_B)(a-b)f'_B(i_B) - k_B z$ and $F_R(i_B) = i_B(1-i_B)(a-b)f'_R(i_B) - k_R z$.*

Proof of Theorem 5 is similar to the proof of Theorem 1 and omitted due to space limitation. Its cyber security implication is: The outcome of playing the Nash equilibrium strategies also depends on the initial system state and the relationship between $k_B$ and $k_R$. As illustrated in Figure 2, if $k_B < k_R$ with $k_R z < \frac{1}{4}(a-b)$, meaning that the attacker is more concerned with its control cost (e.g., reluctant to use/expose its advanced attack tools such as zero-day exploits) than the defender, then $F_B(i_B) = 0$ has two roots $i_1, i_2$ and $F_R(i_B) = 0$ has two roots $i_3, i_4$. Then, we have $i_1 < i_3 < i_4 < i_2$ (the only possibility under the given conditions). Therefore, the outcomes under the Nash equilibrium strategies are summarized as follows:

- If $i_B(0) < i_1$, then $i_B(t) = i_B(0)$ and $i_R(t) = i_R(0)$ for all $t > 0$ because $\hat{\pi}_B = \hat{\pi}_R = 0$ are the Nash equilibrium strategies.
- If $i_3 > i_B(0) > i_1$, then $\hat{\pi}_B = 1$ and $\hat{\pi}_R = 0$ until $i_B = i_3$, which implies that $i_B(t)$ strictly increases until $i_B = i_3$. When $i_B(t) = i_3$ at some point in time $t = t_1$, $\hat{\pi}_B = \hat{\pi}_R = 1$ implies $i_B(t) = i_3$ for $t > t_1$.

**Table 1.** Nash equilibrium strategies for defender and attacker in various cases.

| $k_B$ | $k_R$ | Roots of $F_B(i_B)=0$ | Roots of $F_R(i_B)=0$ | Nash equilibria |
|---|---|---|---|---|
| $k_B z < \frac{1}{4}(a-b)$ | $k_R z < \frac{1}{4}(a-b)$ | $0 < i_1 < i_2 < 1$ | $0 < i_3 < i_4 < 1$ | $\hat{\pi}_B = \begin{cases} 0 & if\ i_B(0) \leq i_1 \\ 1 & if\ i_1 < i_B(0) < i_2 \\ 0 & if\ i_B(0) \geq i_2 \end{cases}$ $\hat{\pi}_R = \begin{cases} 0 & if\ i_B(0) < i_3 \\ 1 & if\ i_3 \leq i_B(0) \leq i_4 \\ 0 & if\ i_B(0) > i_4 \end{cases}$ |
| $k_B z < \frac{1}{4}(a-b)$ | $k_R z = \frac{1}{4}(a-b)$ | $0 < i_1 < i_2 < 1$ | $i_3 = i_4 = \frac{1}{2}$ | $\hat{\pi}_B = \begin{cases} 0 & if\ i_B(0) \leq i_1 \\ 1 & if\ i_1 < i_B(0) < i_2 \\ 0 & if\ i_B(0) \geq i_2 \end{cases}$ $\hat{\pi}_R = \begin{cases} 0 & if\ i_B(0) < i_3 \\ 1 & if\ i_B(0) = i_3 \\ 0 & if\ i_B(0) > i_3 \end{cases}$ |
| $k_B z < \frac{1}{4}(a-b)$ | $k_R z > \frac{1}{4}(a-b)$ | $0 < i_1 < i_2 < 1$ | No real-valued roots | $\hat{\pi}_B = \begin{cases} 0 & if\ i_B(0) \leq i_1 \\ 1 & if\ i_1 < i_B(0) < i_2 \\ 0 & if\ i_B(0) \geq i_2 \end{cases}$ $\hat{\pi}_R = 0$ |
| $k_B z = \frac{1}{4}(a-b)$ | $k_R z < \frac{1}{4}(a-b)$ | $0 < i_1 = i_2 = \frac{1}{2}$ | $0 < i_3 < i_4 < 1$ | $\hat{\pi}_B = \begin{cases} 0 & if\ i_B(0) < i_1 \\ 1 & if\ i_B(0) = i_1 \\ 0 & if\ i_B(0) > i_2 \end{cases}$ $\hat{\pi}_R = \begin{cases} 0 & if\ i_B(0) \leq i_3 \\ 1 & if\ i_3 < i_B(0) < i_4 \\ 0 & if\ i_B(0) \geq i_4 \end{cases}$ |
| $k_B z = \frac{1}{4}(a-b)$ | $k_R z = \frac{1}{4}(a-b)$ | $0 < i_1 = i_2 = \frac{1}{2}$ | $i_3 = i_4 = \frac{1}{2}$ | $\hat{\pi}_B = \begin{cases} 0 & if\ i_B(0) < i_1 \\ \pi_R & if\ i_B(0) = i_1 \\ 0 & if\ i_B(0) > i_2 \end{cases}$ $\hat{\pi}_R = \begin{cases} 0 & if\ i_B(0) < i_3 \\ \pi_B & if\ i_B(0) = i_3 \\ 0 & if\ i_B(0) > i_3 \end{cases}$ |
| $k_B z = \frac{1}{4}(a-b)$ | $k_R z > \frac{1}{4}(a-b)$ | $0 < i_1 = i_2 = \frac{1}{2}$ | No real-valued roots | $\hat{\pi}_B = 0,\ \hat{\pi}_R = 0$ |
| $k_B z > \frac{1}{4}(a-b)$ | $k_R z < \frac{1}{4}(a-b)$ | No real-valued roots | $0 < i_3 < i_4 < 1$ | $\hat{\pi}_B = 0$ $\hat{\pi}_R = \begin{cases} 0 & if\ i_B(0) \leq i_3 \\ 1 & if\ i_3 < i_B(0) < i_4 \\ 0 & if\ i_B(0) \geq i_4 \end{cases}$ |
| $k_B z > \frac{1}{4}(a-b)$ | $k_R z = \frac{1}{4}(a-b)$ | No real-valued roots | $i_3 = i_4 = \frac{1}{2}$ | $\hat{\pi}_B = 0,\ \hat{\pi}_R = 0$ |
| $k_B z > \frac{1}{4}(a-b)$ | $k_R z > \frac{1}{4}(a-b)$ | No real-valued roots | No real-valued roots | $\hat{\pi}_B = 0,\ \hat{\pi}_R = 0$ |

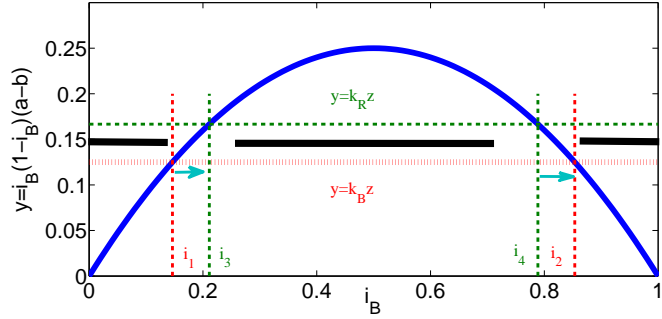

**Fig. 2.** Illustration of the roots of $F_B(i_B) = 0$ with $f_B(i_B) = 1 - i_B$, and the roots of $F_R(i_B) = 0$ with $f_R(i_B) = i_B$, where $a - b = 1$, $k_B z = 1/8$ and $k_R z = 1/6$. The $x$-axis represents $i_B$ and the $y$-axis represents $y(i_B) = i_B(1 - i_B)(a - b)$. Arrows indicate the directions the outcome under the Nash equilibrium heads for. Black-colored bars indicate that the trajectory under the Nash equilibrium stays static.

- If $i_4 > i_B(0) > i_3$, then $i_B(t) = i_B(0)$ and $i_R(t) = i_R(0)$ for all $t > 0$ because $\hat{\pi}_B = \hat{\pi}_R = 1$.
- If $i_2 > i_B(0) > i_4$, then $\hat{\pi}_B = 1$ and $\hat{\pi}_R = 0$ until $i_B = i_2$, which implies that $i_B(t)$ strictly increases until $i_B = i_2$. When $i_B(t) = i_2$ at some point in time $t = t_2$, $\hat{\pi}_B = \hat{\pi}_R = 1$ implies $i_B(t) = i_3$ for $t > t_2$.
- If $i_B(0) > i_2$, then $i_B(t) = i_B(0)$ and $i_R(t) = i_R(0)$ for all $t > 0$ because $\hat{\pi}_B = \hat{\pi}_R = 1$.

If $k_R > \frac{1}{4}(a-b) > k_B$, meaning that the attacker is extremely concerned with its control cost (e.g., not willing to easily use/expose its advanced attack tools such as zero-day exploits) but the defender is not, then it always holds that $\hat{\pi}_R = 0$ because $F_R(i_B) = 0$ has no root but $F_B(i_B) = 0$ has two roots $i_1 < i_2$. From Table 1, we see that $\lim_{t \to \infty} i_B(t) = 1$ always holds, namely that the attacker gives up using its advanced attack tools.

If both $k_B > \frac{1}{4}(a-b)$ and $k_R > \frac{1}{4}(a-b)$, meaning that both the defender and the attacker are extremely concerned with their control costs (i.e., neither the defender wants to easily use/expose its advanced active defense tools, nor the attacker wants to use/expose its advanced attack tools such as zero-day exploits), then it always holds that $\hat{\pi}_B = \hat{\pi}_R = 0$ because $F_B(i_B) = 0$ and $F_R(i_B) = 0$ have no real-valued roots. As a result, $i_B(t) = i_B(0)$ for any $t > 0$.

The scenarios that one or both $F_B(i_B) = 0$ and $F_R(i_B) = 0$ have one root can be regarded as degenerated cases of the above. Moreover, the cases of $k_B > k_R$ (i.e., the defender is more concerned about its control cost, such as not willing to easily use/expose its advanced active defense tools), the outcomes under the Nash equilibria can be derived analogously.

## 6  Conclusion

We have investigated how to optimize active cyber defense, by presenting optimal control solutions for strategic defenders against non-strategic attackers, and identifying Nash equilibrium strategies for strategic defenders and attackers. We have discussed the cyber security implications of the theoretic results.

This paper brings interesting problems for future research. First, it is interesting to extend the models to accommodate nonlinear $f_B(\cdot)$ and $f_R(\cdot)$. Second, the models are geared toward active cyber defense. A comprehensive defense solution, as hinted in our analysis, should require the *optimal* integration of reactive, active, and proactive cyber defense. Therefore, we need to extend the models to accommodate reactive defense and proactive cyber defense. Moreover, it is interesting to investigate how to extend the models to accommodate moving target defense, which has not be systematically evaluated yet [7]. Third, how to extend the models to accommodate the underlying network structures?

**Acknowledgement.** Wenlian Lu was jointly supported by the Marie Curie International Incoming Fellowship from the European Commission (no. FP7-PEOPLE-2011-IIF-302421), the National Natural Sciences Foundation of China (no. 61273309), the Shanghai Guidance of Science and Technology (SGST) (no.

09DZ2272900) and the Laboratory of Mathematics for Nonlinear Science, Fudan University. Shouhuai Xu was supported in part by ARO Grant #W911NF-12-1-0286 and AFOSR Grant FA9550-09-1-0165. Any opinions, findings, and conclusions or recommendations expressed in this material are those of the author(s) and do not necessarily reflect the views of any of the funding agencies.

## A   Proof of Theorem 1

*Proof.* By the Dynamic Programming (DP) argument [3], we know that defender **B**'s value function of the optimal solution can be defined as:

$$\phi(i_B) = \inf_{0 \leq \pi_B(\cdot) \leq 1} \left\{ J_B(\pi_B(\cdot)) = \int_0^\infty e^{-zt}(f_B(i_B(t)) + k_B \cdot \pi_B(t))dt \right\}. \quad (13)$$

This leads to the following Bellman equation:

$$z\phi(i_B) = \inf_{0 \leq \pi_B(\cdot) \leq 1} \left\{ f_B(i_B) + k_B\pi_B(t) + \phi'(i_B)[\alpha_B i_B(1-i_B) - \alpha_R i_B(1-i_B)] \right\}$$
$$= \inf_{0 \leq \pi_B(\cdot) \leq 1} H_B(i_B, \pi_B(t), \phi'(i_B))$$
$$= \inf_{0 \leq \pi_B(\cdot) \leq 1} H_B(i_B, \pi_B(t), p), \; where \; p = \phi'(i_B). \tag{14}$$

From (5), we know that the optimal strategy $\hat{\pi}_B$ takes the form:

$$\hat{\pi}_B = \mathbf{1}_{k_B + pi_B(1-i_B)(a-b)<0} + u_B \mathbf{1}_{k_B + pi_B(1-i_B)(a-b)=0}, \tag{15}$$

where $\mathbf{1}$ is the indicator function. The infimum of Hamiltonian (3) is:

$$\inf_{0 \leq \pi_B(\cdot) \leq 1} H_B(i_B, \pi_B, p)$$
$$= f_B(i_B) + [k_B + pi_B(1-i_B)(a-b)]\mathbf{1}_{k_B + pi_B(1-i_B)(a-b)<0} + p(b-\alpha_R)i_B(1-i_B).$$

Hence, we have

$$z\phi(i_B)$$
$$= f_B(i_B) + [k_B + pi_B(1-i_B)(a-b)]\mathbf{1}_{k_B + pi_B(1-i_B)(a-b)<0} + p(b-\alpha_R)i_B(1-i_B)$$
$$= f_B(i_B) + [k_B + \phi'(i_B)i_B(1-i_B)(a-b)]\mathbf{1}_{k_B + \phi'(i_B)i_B(1-i_B)(a-b)<0} +$$
$$\phi'(i_B)(b-\alpha_R)i_B(1-i_B). \tag{16}$$

Let $y(i_B) = k_B + \phi'(i_B)i_B(1-i_B)(a-b)$. In what follows, we are to verify that (7) satisfies (15) with $\phi(i_B)$ defined by (16), which means that (7) minimizes the Hamiltonian in the term of (13). This completes the proof.

In order to verify that (7) satisfies (15) with $\phi(i_B)$ defined by (16), we differentiate (16) with respect to $i_B$ to obtain

$$[(b-\alpha_R) + (a-b)\mathbf{1}_{y<0}]i_B(1-i_B)y' - zy - F_B(i_B) = 0, \tag{17}$$

which can be rewritten as

$$y' - \frac{z}{[(b-\alpha_R) + (a-b)\mathbf{1}_{y<0}]i_B(1-i_B)}y - \frac{F_B(i_B)}{[(b-\alpha_R) + (a-b)\mathbf{1}_{y<0}]i_B(1-i_B)} = 0. \tag{18}$$

If $k_B + \phi'(i_B)i_B(1-i_B)(a-b) < 0$ namely $y(i_B) < 0$, (18) should be

$$\frac{d}{dx}\left[y(x)e^{-\int_0^x \frac{z}{(a-\alpha_R)\xi(1-\xi)}d\xi}\right] - \frac{F_B(x)}{(a-\alpha_R)x(1-x)}e^{-\int_0^x \frac{z}{(a-\alpha_R)\xi(1-\xi)}d\xi} = 0. \tag{19}$$

If $k_B + \phi'(i_B)i_B(1-i_B)(a-b) > 0$ namely $y(i_B) > 0$, (18) should be:

$$\frac{d}{dx}\left[y(x)e^{\int_x^1 \frac{z}{(b-\alpha_R)\xi(1-\xi)}d\xi}\right] - \frac{F_B(x)}{(b-\alpha_R)x(1-x)}e^{\int_x^1 \frac{z}{(b-\alpha_R)\xi(1-\xi)}d\xi} = 0. \tag{20}$$

Therefore, we only need to prove that the optimal defense strategy (7) satisfies (17), namely (19) or (20). The proof is split into cases, depending on $x$ residing in interval $(i_2, 1)$ or $(0, i_1)$ or $(i_1, i_2)$, or $x = i_1$, or $x = i_2$.

**Case 1: $i_2 < x < 1$.** By (20), we have

$$y(x)e^{\int_x^1 \frac{z}{(b-\alpha_R)\xi(1-\xi)} d\xi} - \int_{i_2}^x \frac{F_B(\zeta)}{(b-\alpha_R)\zeta(1-\zeta)} e^{\int_\zeta^1 \frac{z}{(b-\alpha_R)\xi(1-\xi)} d\xi} d\zeta = 0.$$

Hence, we have

$$y(x) = \int_{i_2}^x \frac{F_B(\zeta)}{(b-\alpha_R)\zeta(1-\zeta)} e^{\int_\zeta^x \frac{z}{(b-\alpha_R)\xi(1-\xi)} d\xi} d\zeta. \qquad (21)$$

Since $i_B > i_2$, we have $y(x) > 0$. Therefore, we have $\hat{\pi}_B = 0$ for $i_B \in (i_2, 1)$.

**Case 2: $0 < x < i_1$.** By (20), we have

$$y(x)e^{-\int_0^x \frac{z}{(b-\alpha_R)\xi(1-\xi)} d\xi} - \int_{i_1}^x \frac{F_B(\zeta)}{(b-\alpha_R)\zeta(1-\zeta)} e^{-\int_0^\zeta \frac{z}{(b-\alpha_R)\xi(1-\xi)} d\xi} d\zeta = 0.$$

Hence,

$$y(x) = \int_0^x \frac{F_B(\zeta)}{(b-\alpha_R)\zeta(1-\zeta)} e^{\int_\zeta^x \frac{z}{(b-\alpha_R)\xi(1-\xi)} d\xi} d\zeta + k_B e^{\int_0^x \frac{z}{(b-\alpha_R)\xi(1-\xi)} d\xi}$$

$$= k_B - \int_0^x \frac{a-b}{b-\alpha_R} f_B'(\zeta) e^{\int_\zeta^x \frac{z}{(b-\alpha_R)\xi(1-\xi)} d\xi} d\zeta \qquad (22)$$

Since $i_B < i_1$, we have $y(x) > 0$. Therefore we have $\hat{\pi}_B = 0$ for $i_B \in (0, i_1)$.

**Case 3: $i_1 < x < i_2$.** By (19), we have

$$y(x)e^{\int_x^1 \frac{z}{(a-\alpha_R)\xi(1-\xi)} d\xi} - \int_{i_2}^x \frac{F_B(\zeta)}{(a-\alpha_R)\zeta(1-\zeta)} e^{\int_\zeta^1 \frac{z}{(a-\alpha_R)\xi(1-\xi)} d\xi} d\zeta = 0.$$

Hence

$$y(x) = \int_{i_2}^x \frac{F_B(\zeta)}{(a-\alpha_R)\zeta(1-\zeta)} e^{\int_\zeta^x \frac{z}{(a-\alpha_R)\xi(1-\xi)} d\xi} d\zeta. \qquad (23)$$

Since $i_B \in (i_1, i_2)$, we have $y(x) < 0$. This implies $\hat{\pi}_B = 1$.

**Cases 4 & 5: $x = i_1$ or $x = i_2$.** By (21,22,23), we have $y(x) = 0$. If $x = i_1$ or $x = i_2$, we can derive $\phi'(i_B) = \frac{-k_B}{i^*(1-i^*)(a-b)}$ from the definition of $y(\cdot)$. According to (16), we have: $z\phi(i^*) = f_B(i^*) + k_B \frac{\alpha_R - b}{a-b}$. Differentiating with respect to $i^*$, we have $-i^*(1-i^*)(a-b)f_B'(i^*) - k_B z = F_B(i^*) = 0$. Consider the singular form $i_B(t) = i^*$ for a period of time. We obtain that $\dot{i}_B \mid_{i_B = i^*} = 0$ and thus $\hat{\pi}_B = u_B = \frac{\alpha_R - b}{a-b}$, where $i^* = i_1$ or $i^* = i_2$. □

## B  Proof of Theorem 3

*Proof.* To solve the minimization problem, we formulate the current value Hamiltonian associated with (9):

$$H_F(i_B, \pi_B, q) = \lambda \pi_B + q(\alpha_B - \alpha_R)i_B(1 - i_B)$$
$$= [\lambda + q(a-b)i_B(1-i_B)]\pi_B + q(b - \alpha_R)i_B(1 - i_B).$$

The adjoint equation is $\dot{q} = -\frac{\partial H_F}{\partial i_B} = -q(\alpha_B - \alpha_R)(1 - 2i_B)$, with the boundary condition

$$H_F(i_B^*(T^*), \pi_B^*(T^*), q(T^*)) + 1 = 0, \tag{24}$$

where $T^*$ denotes the optimal hitting time that $i_B(T^*) = i_e$, $\pi_B^*(\cdot)$ denotes the optimal feedback control, and $i_B^*(\cdot)$ denotes the corresponding trajectory.

The optimal control $\pi_B^*$ is obtained by minimizing Hamiltonian $H_F(i_B, \pi_B, q)$. Since $H_F(i_B, \pi_B, q)$ is linear in $\pi_B$, the optimal control $\pi_B^*$ takes the following bang-bang form:

$$\pi_B^* = \begin{cases} 1 & if\ \frac{\partial H_F}{\partial \pi_B} < 0 \\ u_B^*\ (0 < u_B^* < 1,\ \text{to be determined}) & if\ \frac{\partial H_F}{\partial \pi_F} = 0 \\ 0 & if\ \frac{\partial H_F}{\partial \pi_B} > 0 \end{cases}$$

where $\frac{\partial H_F}{\partial \pi_B} = \lambda + q(a-b)i_B(1-i_B)$. From (24), there are two possibilities: (i). If $\frac{\partial H_F}{\partial \pi_B} \geq 0$, then $0 = \frac{b-\alpha_R}{a-b}(\frac{\partial H_F}{\partial \pi_B} - \lambda) + 1$, which implies $\frac{\partial H_F}{\partial \pi_B} = \frac{a-b}{\alpha_R - b} + \lambda$ is a positive constant. (ii). If $\frac{\partial H_F}{\partial \pi_B} < 0$, then $0 = \frac{\partial H_F}{\partial \pi_B} + \frac{b-\alpha_R}{a-b}(\frac{\partial H_F}{\partial \pi_B} - \lambda) + 1$, which implies $\frac{\partial H_F}{\partial \pi_B} = \frac{a-b}{a-\alpha_R}\left[\frac{b-\alpha_R}{a-b}\lambda - 1\right]$ is a negative constant. It can be seen that only under the above (ii), the constraint $i_B(T) = i_e$ can be obtained for some $T$. Then, the solution to the optimal fast control should be $\pi_B(t) = 1$ for all time. So $(\pi_B^*, T^*) = (1, T_1)$, where $T_1$ satisfies

$$i_B(T_1) = \frac{\frac{i_0}{1-i_0}e^{(a-\alpha_R)T_1}}{1 + \frac{i_0}{1-i_0}e^{(a-\alpha_R)T_1}} = i_e,$$

that is, $T_1 = \frac{1}{a-\alpha_R} \ln\left(\frac{i_e}{1-i_e}\frac{1-i_0}{i_0}\right)$. This completes the proof. □

## C  Proof of Theorem 4

*Proof.* To solve the optimization problem, we formulate the current value Hamiltonian associated with (11):

$$H_F(i_B, \pi_B, q) = \lambda \pi_B^2 + q(\alpha_B - \alpha_R)i_B(1 - i_B)$$
$$= \lambda \pi_B^2 + q(a-b)i_B(1-i_B)\pi_B + q(b - \alpha_R)i_B(1 - i_B).$$

The adjoint equation is $\dot{q} = -\frac{\partial H_F}{\partial i_B} = -q(\alpha_B - \alpha_R)(1 - 2i_B)$, and the boundary condition is
$$H_F(i_B(T^{**}), \pi_B^{**}(T^{**}), q(T^{**})) + 1 = 0, \quad (25)$$
where $T^{**}$ denotes the optimal final time, $\pi_B^{**}(\cdot)$ denotes the optimal feedback control, $i_B(\cdot)$ denotes the corresponding trajectory, and $i_B(T^{**}) = i_e$. Let $D = q(a-b)i_B(T^{**})(1 - i_B(T^{**}))$. From (25) we have
$$H_F(i_B(T^{**}), \pi_B^{**}(T^{**}), q(T^{**})) + 1 = \lambda(\pi_B^{**})^2 + D\pi_B^{**} + \frac{b - \alpha_R}{a - b}D + 1 = 0. \quad (26)$$

The optimal control, $\pi_B^{**}$, is obtained by minimizing the Hamiltonian $H_F(i_B, \pi_B, q)$. Because the Hamiltonian $H_F(i_B, \pi_B, q)$ is quadratic in $\pi_B$, the optimal control, $\pi_B^{**}$, takes the following form:
$$\pi_B^{**} = \begin{cases} 1 & if\ -\frac{D}{2\lambda} > 1 \\ -\frac{D}{2\lambda}\ (0 < u_B^* < 1,\ \text{to be determined}) & if\ 0 \leq -\frac{D}{2\lambda} \leq 1 \\ 0 & if\ -\frac{D}{2\lambda} < 0. \end{cases}$$

From (26), we know there are three possibilities. (i). If $-\frac{D}{2\lambda} < 0$, then $0 = \frac{b-\alpha_R}{a-b}D + 1$, namely that $D = \frac{a-b}{\alpha_R - b}$ is a positive constant. (ii) If $-\frac{D}{2\lambda} > 1$, then $0 = \frac{b-\alpha_R}{a-b}D + 1$, namely that $D = -\frac{a-b}{a-\alpha_R}(\lambda + 1)$ is also a constant. Note that $D < -2\lambda$ if and only if $a - b \leq 2(\alpha_R - b)$, or if and only if $\lambda < \frac{a-b}{a+b-2\alpha_R}$ and $a - b > 2(\alpha_R - b)$. (iii). If $0 \leq -\frac{D}{2\lambda} \leq 1$, then
$$0 = \lambda\left(-\frac{D}{2\lambda}\right)^2 - 2\lambda\left(-\frac{D}{2\lambda}\right)^2 - \frac{b - \alpha_R}{a - b}2\lambda\left(-\frac{D}{2\lambda}\right) + 1,$$
namely that
$$D = 2\frac{b - \alpha_R}{a - b}\lambda - \sqrt{4\left(\frac{b - \alpha_R}{a - b}\lambda\right)^2 + 4\lambda}$$
is a constant. Note that $-\frac{D}{2\lambda} \in (0, 1)$ if and only if $\lambda \geq \frac{a-b}{a+b-2\alpha_R}$ and $a - b > 2(\alpha_R - b)$.

In term of minimizing the Hamiltonian $H_F$ under the above case (i), we have $\pi_B^{**} = 0$ for all time, which is impossible to obtain $i_B(T) = i_e$; under the above case (ii), we have $\pi_B^{**} = 1$ for all time; under the above case (iii), we have $\pi_B^{**} = \frac{D}{-2\lambda}$ for all time. To sum up, we have
$$(\pi_B^{**}, T^{**}) = \begin{cases} (u^*, T_2)\ if\ \lambda \geq \frac{a-b}{a+b-2\alpha_R}\ \text{and}\ a - b > 2(\alpha_R - b) \\ (1, T_3)\ \text{otherwise} \end{cases} \quad (27)$$
where $u^* = \frac{D}{-2\lambda} = \frac{\alpha_R - b}{a - b} + \sqrt{\left(\frac{b-\alpha_R}{a-b}\right)^2 + \frac{1}{\lambda}}$, and $T_2$ and $T_3$ satisfy
$$i_B(T_2) = \frac{\frac{i_0}{1-i_0}e^{(b+(a-b)u^* - \alpha_R)T_2}}{1 + \frac{i_0}{1-i_0}e^{(b+(a-b)u^* - \alpha_R)T_2}} = i_e, \quad i_B(T_3) = \frac{\frac{i_0}{1-i_0}e^{(a - \alpha_R)T_3}}{1 + \frac{i_0}{1-i_0}e^{(a - \alpha_R)T_3}} = i_e,$$
respectively. This completes the proof. □